\documentclass[prb,aps,twocolumn]{revtex4}
\usepackage{amsfonts}
\usepackage{amsmath}
\usepackage{amsbsy}
\usepackage{graphicx}

\renewcommand\vec[1]{\mathbf{#1}}
\newcommand\rvec[1]{\bar{\mathbf{#1}}}

\begin{document}

\title{Electrically controlled  persistent spin currents  at the interface of multiferroic oxides}

\author{Chenglong Jia and  Jamal Berakdar}

\affiliation{Institut f\"ur Physik, Martin-Luther Universit\"at Halle-Wittenberg,
06120 Halle (Saale), Germany}

\begin{abstract}
We predict the appearance of a persistent spin current  in
a two-dimensional electron gas formed at the interface of multiferroic oxides
 with a transverse helical magnetic order. No charge current is generated. This is the result
 of an effective spin-orbit coupling generated
 by the topology of the oxide local magnetic moments.
  The effective coupling  and the generated spin current
   depend linearly on the magnetic spiral  helicity
   which, due to the magneto-electric coupling, is tunable by a transverse electric field
   offering thus a new mean for the generation
   of electrically controlled  persistent spin currents.
\end{abstract}

\maketitle
\section{Introduction}
Nanoscience research is fueled by the spectacular
functionalities that emerge from the
controlled composition of different materials
down to the atomic level. A recent example is
the appearance of a metallic phase
with a high carrier mobility confined to the interface
between insulating  oxides \cite{Oxide0}
such as LaTiO$_3$/SrTiO$_3$   or  LaAlO$_3$/SrTiO$_3$  \cite{Oxide}.
This sheet of two-dimensional electron gas (2DEG) has been
laterally confined and
patterned \cite{Oxide2} to achieve nanometer-sized tunnel junctions and field-effect
transistors \cite{Oxide3}, thus paving the way for oxide-based Nanoelectronics \cite{Oxide2} with
a multitude of technological applications \cite{SFT}.
Further functionalities are expected when utilizing the residual properties of the oxides.
E.g., an important group of Mott insulating oxides such as RMnO$_3$ (R= Tb, Dy, Gd, and Eu$_{1-x}$Y$_x$) \cite{RMnO3} and LiCu$_2$O$_2$ \cite{LiCu2O2}
are multiferroics with a  noncollinear magnetic phase.
The  origin of the spontaneous electric polarization is argued  \cite{KNB} to be the
spin current associated with the spiral magnetic order. As shown experimentally,
 due to the magneto-electric coupling, the  helicity associated with the
  spin spiral structure of the multiferroics is tunable  from clockwise
  to counterclockwise
  type by a small electric field ($\sim 1 kV/cm$) \cite{helicity}.

\begin{figure}[b]
\includegraphics[width=7cm,angle=90]{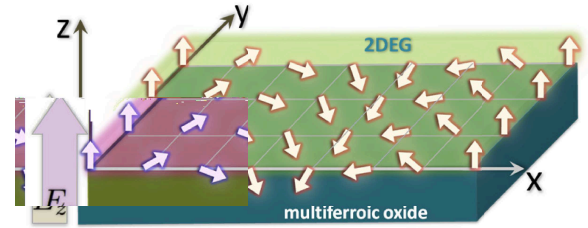}
\caption{(Color online) Schematic of proposed system. The $x$-axis is defined along the direction of spiral ordering, for example, the [110] direction in TbMnO$_3$. The spiral plane of multiferroic (below) is perpendicular to the 2DEG (above).   Due to magneto-electric coupling the spin helicity is controlled by the transverse electric field $E_z$. }
\label{fig::layer}
\end{figure}

   In this paper we show theoretically, that
   a 2DEG formed at the surface of a multiferroic oxide (Fig.\ref{fig::layer})
   such as the $ab$ plane of TbMnO$_3$ \cite{RMnO3}
   experiences an effective spin-orbit interaction (SOI) that linearly depends on the carriers wave vector
   and on the helicity of the oxide's magnetic order and hence is controllable by a lateral electric field.
 As a result an electrically tunable \emph{persistent spin current}
 is shown to build up  in the 2DEG. No charge current is generated.
 The origin of this effect lies in the topological structure of the local magnetic moment at the oxides
 interface.
Spin currents are actively discussed in the field of semiconductor-based
 spintronics \cite{Nagaosa,Berry,semiconductor,intrinsic,extrinsic}. There, SOI plays also a vital role.
  In semiconductors however,  a finite dissipative charge current is also generated by the applied in-plane electric field. Hence, the persistent spin current  in insulator \cite{insulator}, as uncovered here,
  has a decisive advantage, as  compared with metals \cite{metal} and semiconductors \cite{semiconductor} and
  adds a new twist  to oxide electronics.

\section{Theoretical formulation}
As sketched in Fig.\ref{fig::layer}, we consider a 2DEG, as realized in Refs.[\onlinecite{Oxide0,Oxide,Oxide2,Oxide3}] at
the interfaces of oxides layers, however one of the layer should be a spiral multiferroic oxide such as
TbMnO$_3$ \cite{RMnO3} or LiCu$_2$O$_2$ \cite{LiCu2O2}.
The spiral structure defines the  $x-z$ plane, whereas the  2DEG is confined to the $x-y$ plane  (cf. Fig.\ref{fig::layer}). At low temperature, the oxide local spin dynamics  is much slower than the 2DEG carrier dynamics and hence we can treat the oxide local moments as classical and static.
  A carrier in the 2DEG with a
 charge $e$  experiences an effective (real) internal  magnetic field  due to the
magnetic spiral and the embedding medium \cite{swan,wannier,luttinger}
  which results in a nonlocal vector potential $\vec A_{in}$. The effect of $\vec A_{in}$  on the
 charge carriers dynamics is subsidiary
 compared to that of
 the exchange field $J\vec{n}_r$  where  $\vec{n}_r$  is a local unit vector field
 describing the geometry of the localized magnetic moments at the oxides interface and  $J$ is the coupling strength.
The exchange interaction originates from the  Coulomb repulsion on the localized
   moments and from Hund's-rule coupling in the magnetically ordered phase \cite{KNB,Kagome,exchange}.
 Thus, the single particle dynamics in the 2DEG  is governed by the Hamiltonian \cite{SFT}
\begin{equation}
H=h_k+h_J=\frac{1}{2m}\vec{P}^2 + J \vec{n}_r \cdot \boldsymbol{\sigma}
\label{eq:h}\end{equation}
where $m$ is the effective electron mass, $\boldsymbol{\sigma}$ is the vector of the Pauli matrices and $\vec P$ is the momentum operator. $\vec{n}_r$ is given by the local magnetization at the multiferroic surface, i.e.
$\vec{n}_r = ( \sin \theta_r, 0, \cos \theta_r )$
where $\theta_r = \vec{q}_m \cdot \vec{r}$ with $\vec{q}_m = (q,0,0)$ being the spin-wave vector of the spiral.
 $\vec A_{in}$ is not included in Eq.(\ref{eq:h}). \cite{B-field}

Applying the unitary local gauge transformation in the spin space
$U_g = \exp({-i {\theta_r} \sigma_y}/2$),
the spatially non-homogeneous term $h_J$ is transformed into the diagonal term \cite{kor} $\tilde h_J=U_{g}^{\dagger} h_J U_g =J \tilde{\sigma}_z$
 (Hereafter, transformed quantities are marked by a tilde).
Physically, this amounts to a rotation of the local quantization axis to align with $\vec{n}_r$ at each site.
$\sigma_y=\tilde{\sigma}_y$ because  $[U_g, {\sigma}_y]=0$. We find further
  $\tilde{\sigma}_x =(\sigma_x \cos \theta_r -\sigma_z \sin \theta_r)$ and
   $\tilde{\sigma}_z=(\sigma_x \sin \theta_r + \sigma_z \cos \theta_r)$.
   The simplicity of $\tilde h_J$ comes at the price of
    introducing  an additional  gauge field
    $\vec{A}_g = -i \hbar U_g^{\dagger} \nabla_{\vec{r}} U_g $
   in the transformed kinetic energy $\tilde h_k$   \cite{gauge}.
  The gauge field $\vec{A}_g $ depends only
  on the geometry of the local magnetization at the oxide interface.
   As shown below,  $\vec{A}_g $
   acts  as a $q$ and momentum-dependent  effective SOI that can be changed electrically because
     $q$ is tunable by a transverse electric field, as shown in Fig.\ref{fig::layer}.

   For clarity, we introduce the scaled variables (denoted by a bar)
   $\rvec{r}= \vec{r} / a$, $\bar{q} = aq$, and $\rvec{k} = a\vec{k}$ where $a$ is the lattice constant and
   $\vec{k}$ is the crystal momentum.
   The scaled energy $\bar{E}$,  and
   the scaled   exchange energy  $\Delta_m$  read

  \begin{eqnarray}
 && \bar{E}= E/\epsilon_0,
    \; \Delta_m = J /\epsilon_0 \mbox{ with }\; \epsilon_0 =\frac{\hbar^2}{2ma^2}.
  \label{eq:a}\end{eqnarray}
Then the scaled Hamiltonian we find the expression
\begin{equation}
\bar{H}= \left[(i \nabla_{\bar{x}} + \frac{\bar{q}}{2} \tilde{\sigma}_y)^2 + (i \nabla_{\bar{y}}
)^2 \right] + \Delta_m \tilde{\sigma}_z
\label{eq:hs}\end{equation}
For a realistic estimates of the parameters of the 2DEG at oxides interfaces we choose the lattice constant
 $a = 5 {\AA}$.
 %
 %
To our knowledge, the effective mass $m$ of 2DEG at oxide interface is not yet determined.
For insulator however, $m$ is usually quite large, e.g. for SrTiO$_3$  $m$ is $\sim 100$ times larger for
GaAs \cite{Oxide}. Here we choose $m/m_e =10$ with $m_e$ being the free-electron mass, which
 sets the unit of energy to $\epsilon_0 \approx 15 meV$.
 The Hamiltonian (\ref{eq:hs}) we can rewrite in the form
  \cite{note}

\begin{equation}
\bar{H}= \bar{k}_x^2 + \bar{k}_y^2 + \bar{q} \bar{k}_x \tilde{\sigma}_y + \Delta_m \tilde{\sigma}_z.
\label{eq:red}\end{equation}
This relation reveal the existence of a  SOI that depends
 linearly on $\bar{q}$ and  $\rvec{k}$, for the collinear spin phase ($\bar{q} \to 0$) this SOI vanishes.
 The dependence on  $\bar{k}_x$ resembles case of
  a semiconductor 2DEG  in a perpendicular magnetic field   with  the Rashba \cite{Rashba} and Dresselhaus \cite{Dresselhaus} SOI having equal strengths.
In this case, when the  magnetic-field vector potential is taken into account one obtains
 a resonant spin Hall conductance; the spin current is carried by a charge Hall conductivity \cite{R-SC}.
 Such a resonance behavior is  present for  a perpendicular spin polarization.
In our oxide system, however, all averaged values of spin polarization vanishes
 due to a zero average magnetization in the original spin basis.

Explicitly diagonalizing the  Hamiltonian (\ref{eq:red}) we obtain the eigenenergies
\begin{equation}
\bar{E}_{\pm}(\rvec{k}) = \bar{k}_x^2 + \bar{k}_y^2 \pm \sqrt{\Delta_m^2 + (\bar{q}\bar{k}_x)^2}
\label{eigenenergies}
\end{equation}
with the eigenstates
\begin{eqnarray}
\label{eigenstates}
|\psi_{+} \rangle = e^{-i \rvec{k} \cdot  \rvec{r}} \left( \begin{array}{cc}  \cos \frac{\phi}{2} \\ i \sin \frac{\phi}{2} \end{array} \right),\;
|\psi_{-} \rangle = e^{-i \rvec{k} \cdot  \rvec{r}}  \left( \begin{array}{cc}  i \sin \frac{\phi}{2} \\ \cos \frac{\phi}{2} \end{array} \right)\end{eqnarray}
where
\begin{eqnarray}
\tan \phi=  \frac{\bar{q}\bar{k}_x}{\Delta_m},\; \; \cos \phi = \frac{\Delta_m }{\sqrt{\Delta_m^2 + (\bar{q}\bar{k}_x)^2}}.
\end{eqnarray}
\begin{figure}[t]
\includegraphics[width=6cm,angle=-90]{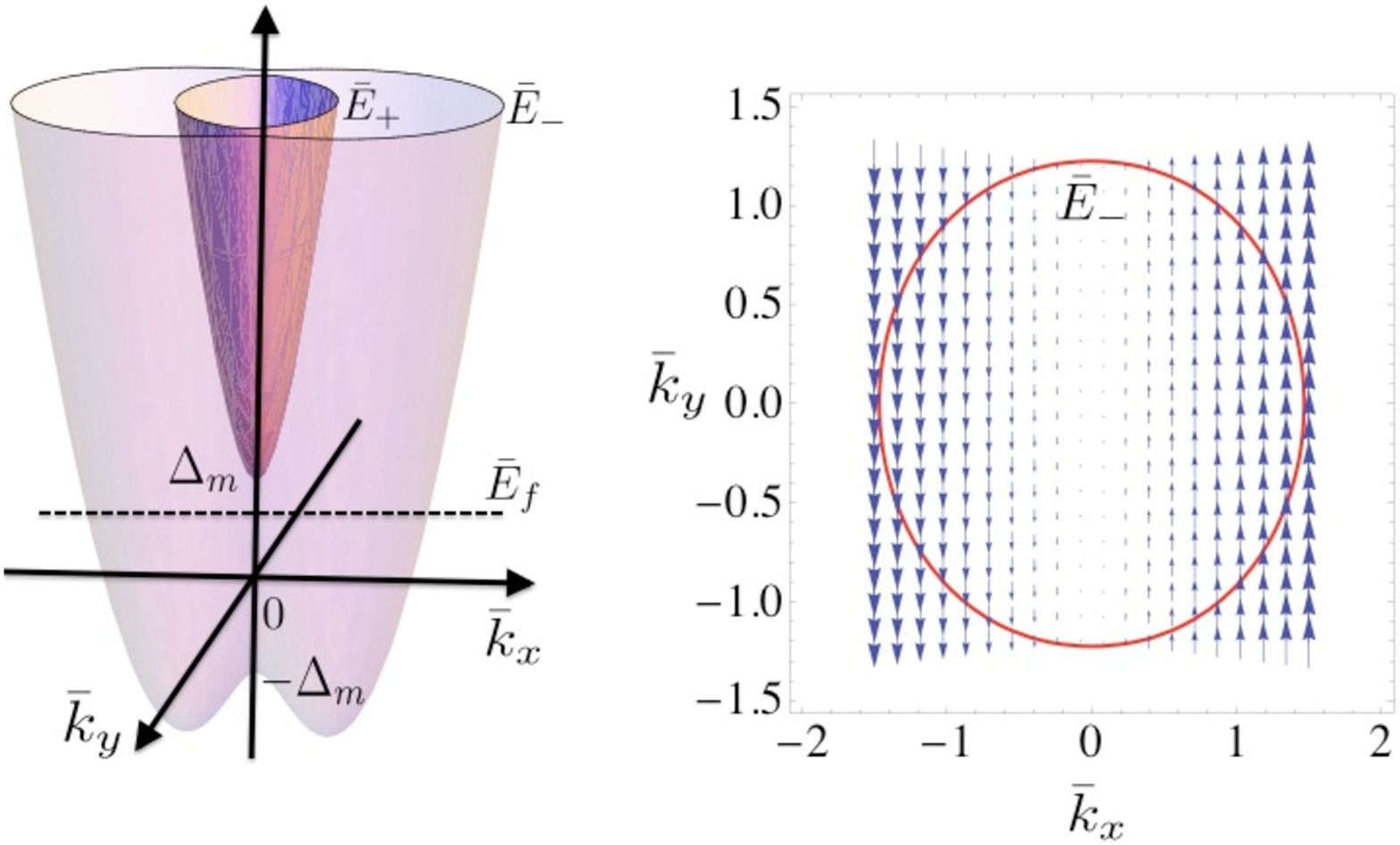}
\caption{(Color online) Energy bands: $\bar{E}_{\pm}$ corresponds to the two energy branches, respectively. When $\bar{E}_f < \Delta_m$, only the low energy band $\bar{E}_{-}$ is involved in the Fermi contour. The arrows represent the effective SOI, $\bar{q}\bar{k}_{x}\tilde{\sigma}_y$. The strengths of the carrier coupling to the local magnetic order is chosen as $\bar{E}_f/ \Delta_m =1/2$ and the spiral wave vector is $\bar{q}= 2\pi /7$.}
\label{fig::energy}
\end{figure}
Due to the effective spin-orbit coupling, the Fermi contours are not parabolic but anisotropic having  $\hat{x}$ and $\hat{y}$ as the symmetry axes, as depicted in Fig.\ref{fig::energy}. Although the spin states in Eq.(\ref{eigenstates}) are not independent of $\rvec{k}$, we still have a disappearance of the Berry phase just as the case without magnetic field in Ref.[\onlinecite{Shen}], which implies that a spin current along spin $\hat{z}$ direction does not exist in the absence and presence of an electric field.
\begin{figure}[t]
\includegraphics[width=6cm,angle=-90]{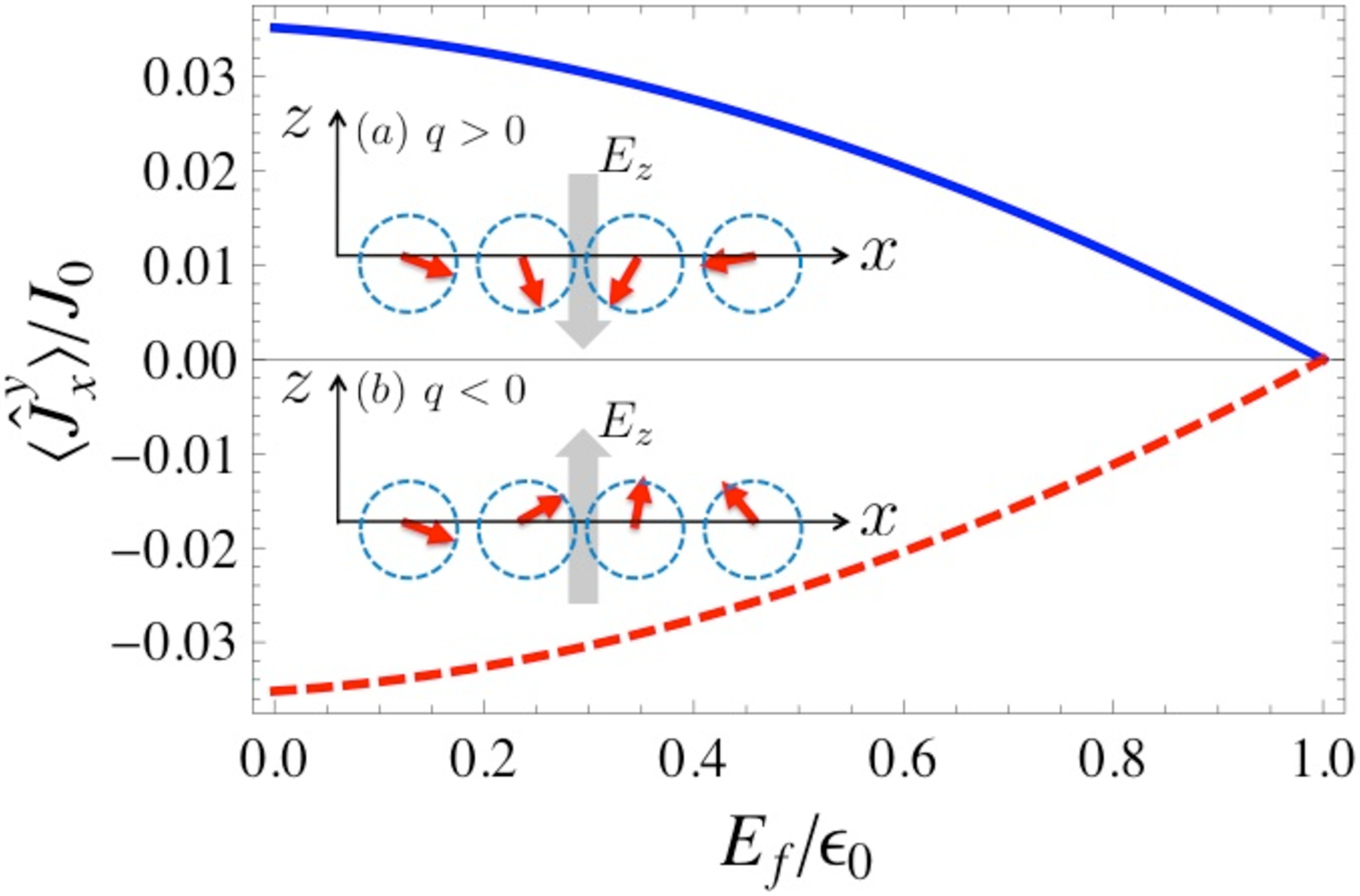}
\caption{The persistent spin current as a function of $E_f$
for positive (a) and negative (b)  helicities when Fermi level intersects only the low energy band. The parameters are chosen
  as $\epsilon_0 = 15 meV$, $J_0 = a\epsilon_0 /2$, $\Delta_m = \epsilon_0$ and $\bar{q} = 2\pi/7$. As
  illustrated in the insets, depending on the direction of an applied
   transverse electric field $E_z$
  the spiral helicity and hence the spin current directions are reversible due to the
  magneto-electric coupling.}
\label{fig::current}
\end{figure}

\section{Persistent spin current}
The expectation value of spin polarizations per electron evaluated
using the eigenstates  (\ref{eigenstates})
are
\begin{eqnarray}
\langle \tilde{\sigma}_y^s \rangle = {s\bar{q}\bar{k}_x \over \sqrt{\Delta_m^2 + (\bar{q}\bar{k}_x)^2} } \\
\langle \tilde{\sigma}_z^s \rangle  =
{s\Delta_m \over \sqrt{\Delta_m^2 + (\bar{q}\bar{k}_x)^2}}
\end{eqnarray}
Here $s=\pm$, the double sign corresponds to the  two branches of the energy dispersion  Eq.(\ref{eigenenergies}).
In the original spin space, $\langle \tilde{\sigma}_z^{s}\rangle$ corresponds spiral spin ordering induced by the exchange interaction between the 2DEG and the local magnetic moments at the oxide surface. Obviously, the $\hat{y}$ spin-polarization component $\langle \sigma_y^{s}\rangle$ is odd in $\bar{k}_x$
 and it vanishes upon summing over all occupied states.
The spin current in the $\hat{x}$ direction is however generally \emph{finite} when the Fermi level intersects only one of the two bands.
To prove this we consider the spin current operator,  defined as
\begin{equation}
\hat{J}_j^i = {\hbar \over 4} ( \sigma_i v_j + v_j \sigma_i)
\end{equation}
where the velocity operators at each $\rvec{k}$ are given by
$ v_x = \partial {H} / \partial p_x  = \frac{\hbar}{2ma}(2\bar{k}_x + \bar{q} \tilde{\sigma}_y )$ and
$ v_y = \partial {H} / \partial p_y = \frac{\hbar \bar{k}_y}{ma}$.
Considering the symmetry $\bar{E}_s(\rvec{k}) = \bar{E}_s(-\rvec{k})$ of the eigenenergies (Eq.\ref{eigenenergies}),
it follows that only $\langle \hat{J}_x^y \rangle $ is finite and is determined by (in unit of $J_0 = \frac{\hbar^2}{4ma}$)
\begin{eqnarray}
\langle \hat{J}_x^y \rangle  &=& \sum_{s=\pm} \int  \frac{d^2 \rvec{k}}{(2 \pi)^2} ( 2 \bar{k}_x \tilde{\sigma}_{y}^s + \bar{q}) f[\bar{E}_s(\rvec{k})]
\end{eqnarray}
where $f[\bar{E}_s(\rvec{k})]$ is the Dirac-Fermi distribution, and $\rvec{k}$ runs all occupied states.
Introducing the wave-vector  parameterization  $\rvec{k} = \bar{k} (\cos \varphi, \sin \varphi)$, it can be analytically shown that $\langle \hat{J}_x^y \rangle =0$ when $\bar{E}_f \ge \Delta_m$. However, as shown in Fig.\ref{fig::current}, the spin current is finite when only the low energy band is intersected by the Fermi level. Interestingly, the spin current  is related
 to the electron density, $n_f$ through the Fermi energy $E_f$. More important,
  the key factor is the \emph{odd} relationship between the spin current and the geometrical spiral structure of the  magnetic ordering, being
   clockwise ($\bar{q}<0$) or anticlockwise ($\bar{q}>0$) (Fig.\ref{fig::current}).
   This is insofar important, as  spin-polarized neutron scattering experiments \cite{helicity}
   on multiferroics evidently
    show  that the helicity of the spiral magnetic order is controllable  by a small ($\sim 1 kV/cm$)
    transverse  electric field, as illustrated in Fig.\ref{fig::current}.

In the absence of an electric field, when the exchange interaction is strong enough, i.e. for large
 $\Delta_m$, the spins of the conduction electrons are initially aligned locally
 parallel to $\vec{n}_r$ at each site, $\langle \tilde{\sigma}_z \rangle = 1$.  Using the Heisenberg equation of the electron-spin  motion, \cite{G-Force} we
   find in the linear response regime that  the 2DEG
   develops a uniform spin polarization
\begin{equation}
\langle \tilde{\sigma}_x \rangle =-\bar{q} \frac{eaE_x}{2\epsilon_0 \Delta_m^2}
\end{equation}
   when an  external electric field is applied along $\hat{x}$ direction. Transforming back demonstrates that the solution for $\langle \tilde{\sigma}_x \rangle$  corresponds to emergence
     of a spiral spin-density wave in the 2DEG   rotating
      in the $x-z$ plane. The direction of the spin polarization is
      orthogonal to the oxide local magnetic moment.
      Furthermore, the linear dependence on $q$  of $\langle \tilde{\sigma}_x \rangle$  allows
       for an electric-field control of the induced spin helicity.

\section{Hall conductivity}  The oxide magnetic order
 is usually not exactly coplanar in the $y-z$ plane but it has a small deviation. Here we simulate this non-coplanar modulation with a slowly varying spiral order with a spin helicity given by $(0,\beta \bar{q},0)$ ($\beta \ll 1$)  along $\hat{y}$ direction. This results in an another
 effective spin-orbit coupling term $\sim \beta \bar{q} \bar{k}_y \tilde{\sigma}_x$
 with a strength $\beta \bar{q}$. In analogy to the semiconductor case,
  this amounts  to the Rashba  and the Dresselhaus SOIs having  different strengths.
   Therefore, we expect in our case the existence of a Hall effect.
   To see this, we diagonalize  the  resulting total Hamiltonian using the  transformation

\begin{eqnarray}
T =  \left( \begin{array}{cc}
\bar{q} \sin {\phi' \over 2} (\frac{-\beta \bar{k}_y + i \bar{k}_x} {F_{\bar{k}}}) & \cos {\phi' \over 2} \\ \cos {\phi' \over 2} & \bar{q} \sin {\phi' \over 2} (\frac{\beta \bar{k}_y + i \bar{k}_x}  {F_{\bar{k}}}) \end{array} \right )
\end{eqnarray}
where
\begin{eqnarray}
 \cos \phi' = \frac{\Delta_m}{\sqrt{\Delta_m^2 + F_{\bar{k}}^2}},\quad
 F_{\bar{k}} = \bar{q} \sqrt{\bar{k}_x^2 + \beta^2 \bar{k}_y^2}.
\end{eqnarray}
The Hall effect in the 2DEG is related to the  nontrivial topology of
 the resulting eigenstates $|\bar k\rangle$ in the momentum space \cite{Berry,haladane,luttinger},
 expressed through the
gauge connection ${\cal A}_{\bar k}=-i\langle \bar k|\nabla_{\bar k}|\bar k\rangle$.
 The  off-diagonal  Hall conductivity is related to  Berry's curvature \cite{Berry}
 $\Omega_{s}^z=\nabla_{\bar k}\times {\cal A}_{\bar k}$
  pointing along the $\hat{z}$ axis, for which  we obtain
\begin{equation}
\Omega_{s}^z = - {s \over 2} \frac{\beta}{\cos^2 \varphi + \beta^2 \sin^2 \varphi} ~ {1 \over \bar{k}} \frac{\partial \cos \phi'}{\partial \bar{k}}.
\end{equation}
$\Omega_{s}^z$  diverges along the $\hat{y}$ axis at very small $\beta$, and  is singular at the origin $\bar{k}=0$. The geometrical Berry phase factor $\gamma_s$ is given by the integral of the curvature over all wave vector
\begin{eqnarray}
\gamma_s &=& \int \Omega_s^z d^2\rvec{k} = s\pi (1- I(\beta,\bar{k}_{s}^{f})), \\
I(\beta,\bar{k}_{s}^{f}) &=& {1 \over {2\pi}} \int_0^{2\pi}  \frac{\beta}{\cos^2 \varphi + \beta^2 \sin^2 \varphi} \frac{\Delta_m}{\sqrt{\Delta_m^2 + F_{\bar{k}_s^{f}}^2}}. \nonumber \\
\end{eqnarray}
The Fermi wave vector $\bar{k}_s^{f}$ is given by the Fermi energy $\bar{E}_f = \bar{k}^2 \pm \sqrt{\Delta_m^2 + F_{\bar{k}}^2}$. At zero temperature, the off-diagonal  Hall conductivity $\sigma_{xy}$ for a full band is equal to the integral over the Brillouin zone of the component of the Berry curvature parallel to $\hat{z}$ and is thus proportional to the Berry phase \cite{luttinger,Berry}, i.e.

\begin{eqnarray}
\sigma_{xy}^s = {e^2 \over \hbar} \int \Omega_s^z {d^2\rvec{k} \over {(2\pi)^2}} = s{e^2 \over {2h}}(1- I(\beta,\bar{k}_{s}^{ f}))
\end{eqnarray}
For a quit small $\beta$, $I(\beta,\bar{k}_{s}^{f}) \rightarrow 0$, and $\sigma_{xy} = -\frac{e^2}{2h}$ is quantized when the only one of two bands is intersected by the Fermi level. Generally, $\sigma_{xy}$ is not quantized, but the transverse conductivity should still be observable.

\section{Summarizing}  A persistent spin current emerges in 2DEG at the interface
 of a helimagnet  due to the spiral geometry of the local magnetic order.
 The spin current is an odd function of the spin helicity and hence electrically
 controllable by a small transverse  electric field that reverse
the spin helicity, making thus a link between spintronics and oxide electronics.
For an in-plane  electronic field along the spiral we
we predict the buildup of carrier spiral spin density wave. The spin Berry phase induced by a chiral magnetic texture in a Kagom\'e lattice has been discussed in Refs.[\onlinecite{Kagome}]. Due to a nonzero spin chirality defined as the mixed product of three spins on a certain plaquette, $\chi_{ijk} = \vec{S}_i \cdot (\vec{S}_j \times \vec{S}_k)$, they showed that the Berry phase contribution to the Hall conductivity is quantized for some values of the band
filling. We also calculated the
Berry curvature and obtained a finite Hall conductivity for even a small derivation from the coplanar oxide helical magnetic order. The transverse conductivity is determined by the chirality ($\beta$), the electron density ($\bar{k}_{s}^{f}$), and the strength of the exchange interaction ($\Delta_m$), $\sigma_{xy}$ can thus be quantized, or posseses a nonmonotonic behavior upon varying these dependent parameters.

This research is supported by the DFG (Germany)
through the project-B7-  in the SFB762:{\it functionality of oxide interfaces}.


\begin{thebibliography}{99}
\bibitem{B-field} Generally the determination of $\vec A_{in}$ is quite involved and is geometry dependent.
 For an order of magnitude estimate  we consider the contribution from the local magnetic moments\cite{Peeters}. The $z$-component of the resulting $\vec{H}$-field reads
%
$$ H_z(r) = \frac{1}{4 \pi} \int d\vec{r}' \frac{\rho_m(r') ( z-z')}{|\vec{r}-\vec{r}'|^3}
 \approx  H_0 \cos qx $$
%
$$ H_0 = -\frac{1}{4 \pi}
 \int d\vec{r}' \frac{z' \rho_m(r')}{|\vec{r}'|^3}, ~~~\rho_{m} (r) = - \nabla \cdot \vec{M}_{in}(r), $$
%
where $\vec{M}_{in}$ is the oxide local magnetization, $\vec{M}_{in} = (-1)^n M_0 \vec{n}_r$.
We thus have $$B_z(r) = \mu_0 \left[H_z(r) + M_z(r)\right]= B_0 \cos qx .$$
%
In the Landau gauge the vector potential is space periodic, i.e.
$\vec{A}_{in} = B_0 (0, {\sin qx \over q}, 0)$.   $\vec{A}_{in}$ induces a periodic potential along the $\hat{x}$ direction with an amplitude  ${eB_0 a^2}\over {\hbar \bar{q}}$ (a is the lattice constant),
which is tiny in our case
(the local magnetic moment is $\approx 4 \mu_B$ on each Mn site in RMnO$_3$ \cite{RMnO3}, the magnetic field is estimated as $B_0 \sim 0.3 T$).
 The  energy associated
with the generated harmonic potential  is  $\hbar e|B_z|/m = |B_z| \frac{m_e}{m} \times 10^{-1} meV$ with $m_e$ being the free-electron mass. The effective mass $m$ for the 2DEG at the oxide interfaces is $m_e / m \sim 1/10$, meaning a  small energy contribution from the vector potential.
\bibitem{note} A uniform energy displacement $\Delta E = \bar{q}^2/4$ is disregarded in $\bar{H}$.
\end{thebibliography}
\end{document}